\documentclass[a4paper,11pt]{article}
\pdfoutput=1 

\usepackage{jinstpub} 

\usepackage{lineno}

\title{\boldmath Validation of high voltage power supplies for the 1-inch photomultipliers of AugerPrime, the Pierre Auger Observatory upgrade}

\author[a,b,1]{G. A. Anastasi,\note{Corresponding author.}}
\author[c,e]{M. Buscemi,}
\author[a,b]{M. Aglietta,}
\author[d,e]{R. Caruso,}
\author[a,b]{A. Castellina,}
\author[d,e]{S. Costa,}
\author[b]{S. Gallian,}
\author[a,b]{A. Gorgi,}
\author[d,e]{N. Guardone,}
\author[d,e]{C. Lombardo,}
\author[b]{R. Wheadon,}
\author[b]{A. Zampieri}

\affiliation[a]{Osservatorio Astrofisico di Torino (INAF), Via Osservatorio 20, Pino Torinese, Torino, Italy}
\affiliation[b]{Istituto Nazionale di Fisica Nucleare (INFN), Sezione di Torino, Via P. Giuria 1, Torino, Italy}
\affiliation[c]{Dipartimento di Fisica e Chimica "E. Segr\`{e}", Universit\`{a} degli Studi di Palermo, Viale delle Scienze Ed. 18, Palermo, Italy}
\affiliation[d]{Dipartimento di Fisica e Astronomia "E. Majorana", Universit\`{a} degli Studi di Catania, Via S. Sofia 64, Catania, Italy}
\affiliation[e]{Istituto Nazionale di Fisica Nucleare (INFN), Sezione di Catania, Via S. Sofia 64, Catania, Italy}

\emailAdd{gioacchino.alex.anastasi@to.infn.it}

\abstract{
In the framework of the upgrade of the Pierre Auger Observatory, a new high voltage module is being employed for the power supply of the 1-inch photomultiplier added to each water-Cherenkov detector of the surface array with the aim of increasing the dynamic range of the measurements.
This module is located in a dedicated box near the electronics and comprises a low consumption DC-DC converter hosted inside an aluminum box.
All the modules have undergone specific tests to verify their reliability in the extreme environmental conditions of the Argentinian pampa.
In this paper, we describe the validation procedure and the facility developed to this aim. The successful results of the tests on the HVPS modules are presented and discussed.
}


\begin{document}
\maketitle
\flushbottom
\section{Introduction}

The Pierre Auger Observatory \cite{auger} is the largest observatory in the world for the study of ultra-high energy (E~$\ge10^{17}$~eV) cosmic rays and is located in the Southern Hemisphere in the province of Mendoza, Argentina.
It is characterized by a hybrid design, employing both a surface detector and a fluorescence detector.
The surface detector consists of an array of water-Cherenkov detectors (WCD) in a triangular grid with 1500 m spacing, and covering about 3000 km$^2$; this area is overlooked by 27 fluorescence telescopes distributed in four sites at the border of the array.

The Pierre Auger Observatory is currently undergoing an important upgrade \cite{augerprime, augerprime-SPMT}, with the main aim of improving the sensitivity to the chemical composition of ultra-high energy cosmic rays.
A scintillator detector is being installed on top of each of the existing WCDs, to disentangle the muonic and electromagnetic components of the shower signal \cite{augerprime-SSD}. Also, a radio detector antenna is being deployed at each station of the surface array to enhance the sensitivity to mass composition for inclined showers \cite{augerprime-radio}. An underground muon detector \cite{augerprime-UMD} is buried under the WCDs in a limited area of the array, to provide a direct measurement of the muonic component.
New electronics (Upgraded Unified Board, UUB) have been developed to process the output of all the existing and new detectors, ensuring improved performances and increasing the acquisition time resolution. 
To extend the dynamic range of the measurement, each WCD will include a small photomultiplier (SPMT), thus allowing to avoid the saturation in the proximity of the shower core \cite{augerprime-UUB}.

Each WCD is equipped with three 9-inches large photo-multipliers (LPMTs) supplied through an active base soldered to the PMT leads and insulated by silicon potting. The base includes the High Voltage (HV) resistive divider, a HV DC-DC converter module and a charge amplifier for the dynode readout \cite{auger}.
The large thermal variation in the Argentinian pampa and the high level of moisture inside the WCD cause malfunctioning mainly related to the HV system, with a rate of about one hundred failures per year over a total of $\sim$5000 LPMTs.
Since the power supply is integrated in the base, any repair operation also requires the substitution of the photomultiplier.
On the other hand, an external HV module allows to minimize the number of failures and to make the maintenance easier. Such a solution is employed for the SPMT, a Hamamatsu R8619 \cite{SPMT-hamamatsu}, by designing a passive base and moving the HV power supply into a separate module located in a dedicated box near the electronics.

In this paper, we define as HVPS module the combination of a single HV channel with a simple interface circuit designed to allow a safe \textit{plug and play} connection to the UUB controlling the activity of each surface detector.
The HVPS module not only supplies the SPMT but also transmits parameters related to temperature and current to the UUB.
A CAEN-A7501 HV DC-DC converter hosted in a metallic box (see sect.~\ref{sec:specifications}) has been chosen for this purpose, characterized by slightly better thermal stability, operational temperature range and power consumption compared to the module used for the LPMTs (SensTech-PS2010/12).
A total of 1650 HVPSs have been validated using a test facility developed at INFN-Torino and an identical one reproduced at INFN-Catania, and are now being deployed at the Pierre Auger Observatory site.
This number is sufficient to instrument the 1493 WCDs internal to the experimental area (that is, excluding the most external border\footnote{Indeed, the highest particle density measurable by the SPMT is collected in the station nearest to the shower core, which due to our trigger
conditions is required to be internal, i.e.~fully surrounded by at least one crown of WCDs.}), leaving a safe number of spare units.
The results for the 90 pre-production modules, characterized by a slightly different design, are excluded in the following.

\section{Specification for the High Voltage Power supplies}
\label{sec:specifications}

The requirements for the high voltage modules supplying the SPMTs were kept the same as for the DC-DC converters mounted on the LPMT bases \cite{paper_bases} which have been successfully operational in the experimental site for more than 15 years.
Such high reliability constraints are motivated by the stressful environmental conditions where they operate, such as: (i) temperature ranging between -10$^\circ$C and 50$^\circ$C and daily variation frequently exceeding 30$^\circ$C; (ii) electric power provided by solar panels, thus requiring to keep the power consumption of the HVPS modules as low as possible.
%
%
Moreover, all the photomultipliers serving a WCD are installed such that the surface of the photocathode is in contact with the plastic window dome viewing the 12~tons water volume from above, so a positive HV supply is preferred (in particular, about 1000~V for the SPMT).

The specifications, reported in detail in table~\ref{table1}, should guarantee optimal functionality for the foreseen lifetime of the detector (more than 10 years).

\begin{figure}[!tbp]
\centering 
\includegraphics[width=.99\textwidth]{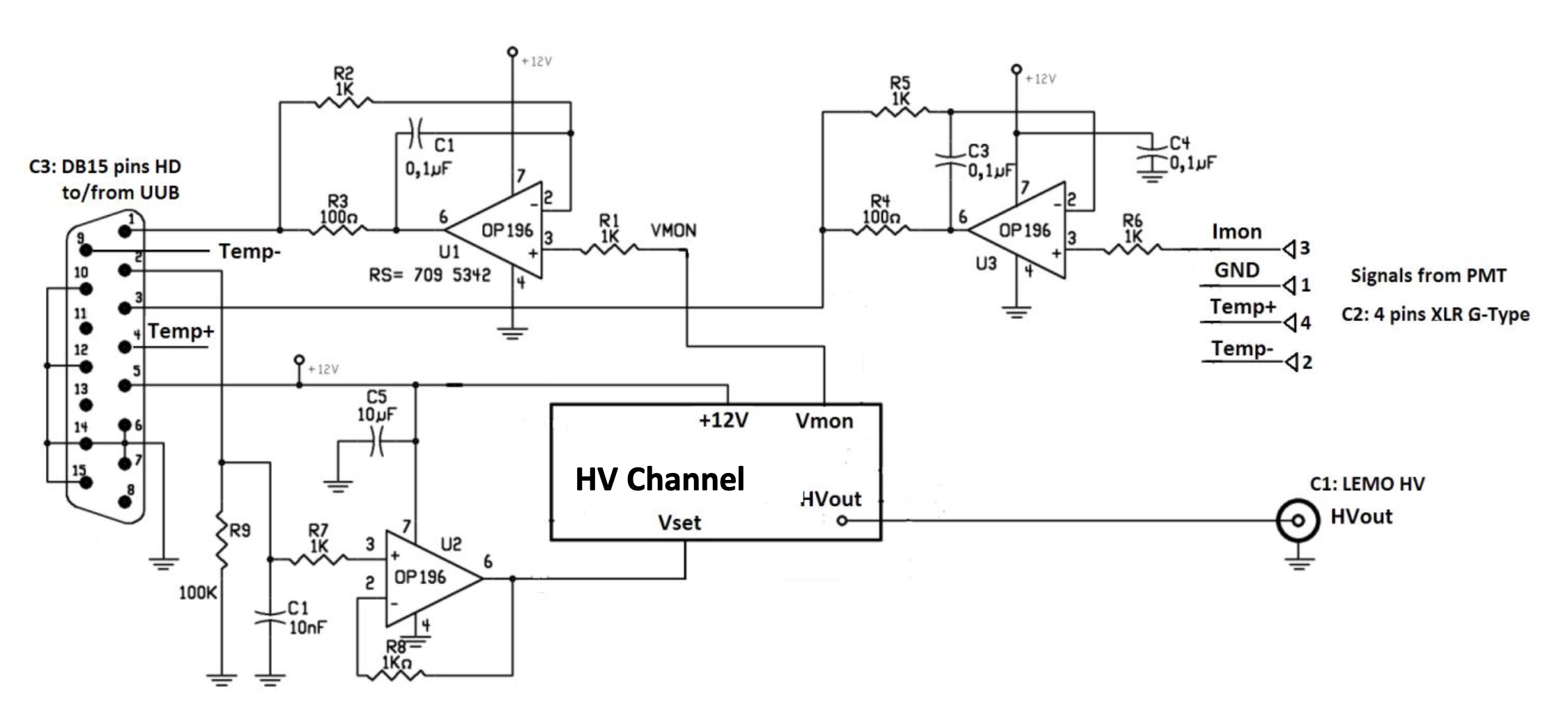}
\caption{\label{fig:schematic_HVPS} A schematics of the interface circuit for the Upgraded Unified Board (UUB).}
\end{figure}

A new single channel HV power supply (A7501B, \cite{CAEN1}) was developed by CAEN -- in a cooperative effort with INFN-Torino -- specifically for large area experiments operating in challenging conditions where the possibility of a prompt maintenance intervention is reduced.
It is composed by a HV DC-DC converter (CAEN-A7501, \cite{CAEN2}) hosted inside an aluminum box designed to match the front-end of the UUB (figure~\ref{fig:schematic_HVPS}), thus equipped with: a LEMO HV connector to deliver the HV output to the PMT; a 4-pin Amphenol G.Type Miniature XLR male connector to monitor the PMT parameters (current and temperature); a high density 15 pin Cannon DB type male connector for the control and 12~V supply by the UUB.


The final design and choice of the components is a balance between performance and cost, which includes the request to the vendor to individually test each HVPS module at room temperature and provide a detailed documentation of the measured parameters. An extract of a typical report can be found in appendix~\ref{appendix:a}.

\begin{table}[!t]
	\centering
    \caption{\label{table1} Required specifications for the HVPS modules.}
    \smallskip
    \renewcommand{\arraystretch}{1.2}

	\begin{tabular}{|p{38mm}|p{62mm}|p{33mm}|}
		\hline
		\textbf{Parameter} & \textbf{Description} & \textbf{Requirement} \\
		\hline
		$\mathrm{V_{cc}}$ & main voltage supply & 12 V $\pm$ 5\%\\
		\hline
		$\Delta$T & temperature range & -20 ~--~ 65 $^\circ$C\\
		\hline
		$\mathrm{HV_{out}}$ & output voltage & 0 -- 2100 V\\
		\hline
		$\mathrm{HV_{max}}$ & maximum output voltage & 2100 V $\pm$ 5\%\\
		\hline
		$\mathrm{I_{out}}$ & output current & 100 $\mu$A @ 20 M$\Omega$\\
		\hline
		$\mathrm{V_{set}}$ & external voltage setting & 0 -- 2.5 V\\
		\hline
		$\mathrm{V_{mon}}$ & output voltage monitor & 0 -- 5 V\\
		\hline
		$\mathrm{I_{mon}}$ & output current monitor & 0 -- 2.5 V\\
		\hline
		$\Delta \mathrm{HV_{max}}/\mathrm{HV_{max}}$ & stability $\mathrm{HV_{max}}$ vs $\mathrm{V_{cc}}$ & < 1\textperthousand\\
		\hline
		$\mathrm{P_{abs}}$ & power absorption & < 500 mW @ $\mathrm{HV_{max}}$\\
		\hline
		 & ripple voltage & < $2 \times 10^{-5}$ @ 20 M$\Omega$\\
		\hline
		$\Delta \mathrm{HV_{max}}/(\mathrm{HV_{max}} \times \Delta \mathrm{T})$ & thermal stability & <$10^{-4}$/$^\circ$C\\
		\hline
		$\mathrm{NL_{HV_{out}}}$ & integral non linearity for $\mathrm{HV_{out}}$ vs $\mathrm{V_{set}}$ for $\mathrm{HV_{out}} > 500$~V & < $2 \times 10^{-3}$\\
        \hline
		$\mathrm{NL_{V_{mon}}}$ & integral non linearity for $\mathrm{V_{mon}}$ vs $\mathrm{HV_{out}}$ for $\mathrm{HV_{out}} > 500$~V & < $2 \times 10^{-3}$\\
		\hline
		 & uniformity of $\mathrm{HV_{max}}$ & < $10^{-3}$\\
		\hline
	\end{tabular}
\end{table}

\section{The validation procedure and the test facility}
\label{sec:facility}
A four step procedure has been developed to perform a complete characterization and validate each HVPS according to the requests listed in table~\ref{table1}. 

At first, the unplugged modules undergo a 4 day long \textit{thermal stress} with controlled temperature cycles between -20$^\circ$C and +65$^\circ$C inside a climatic chamber, a test useful to check the reliability of the HVPS assembly and soldering.
Two further steps (from now on indicated as \textit{Linearity} and \textit{Uniformity} tests) are then performed at room temperature. The latter test consists of a 24-hour \textit{thermal stress} during which the HVPSs are turned on.
All the HVPSs have been measured in their full operational range and with the requested accuracy using the test bench and the procedures described in the following. 
The vendor agreed to trust this evaluation in the definition of accepted or rejected (i.e.~outside the required specifications) units.
A module that fails to satisfy one of the requirements is tested again and, if the failure persists, is eventually rejected.\\
In addition to the described procedure, two tests are performed at random on $\sim$20\% of the modules:
\begin{itemize}
\vspace{-0.1cm}
    \item the HVPS is short-circuited to verify the protection and the subsequent operability;
\vspace{-0.1cm}
    \item the ripple on the high voltage output is checked at $\mathrm{V_{set}}=2.5$~V with a digital oscilloscope.
\vspace{-0.1cm}
\end{itemize}
In all cases the short-circuit protection worked fine.
Furthermore, the ripple amplitude results on average $2 \times 10^{-6}$ @ 20~M$\Omega$, that is a factor $\sim$10 lower with respect to the requirement in table~\ref{table1} and to the measured value on the SensTech DC-DC converter mounted on the LPMT base.

\subsection{Linearity measurements}
\label{subsec:linearity}

During the linearity test, a complete study of the HV module response for increasing values of the $\mathrm{V_{set}}$ parameter is performed. The control voltage $\mathrm{V_{set}}$ is increased from 0 V up to its maximum value of 2.5~V (corresponding to HV$_{\mathrm{out}}$ = HV$_{\mathrm{max}}$ $\sim$ 2100 V), following a ramp up with ten steps of 0.25~V. For each step the following monitoring parameters are registered and written in a file to be analysed offline : 
\begin{itemize}
\vspace{-0.1cm}
    \item output voltage monitor $\mathrm{V_{mon}}$, which is proportional to $\mathrm{HV_{out}}$;
\vspace{-0.1cm}
    \item output current monitor $\mathrm{I_{mon}}$, proportional to the voltage drop across a 20~k$\Omega$ resistor in series with the 20~M$\Omega$ load used to mimic the SPMT; in these conditions $\mathrm{I_{mon}}$ is equivalent to the value of $\mathrm{HV_{out}}$ reduced by a factor $\sim$1000;
\vspace{-0.1cm}
    \item absorption current $\mathrm{I_{V_{cc}}}$, that for a supply voltage $\mathrm{V_{cc}} = 12$~V must result in a power absorption lower than 500~mW.
\vspace{-0.1cm}
\end{itemize}

\begin{figure}[tb]
\centering
\hspace{-.75cm}
\includegraphics[width=.49\textwidth]{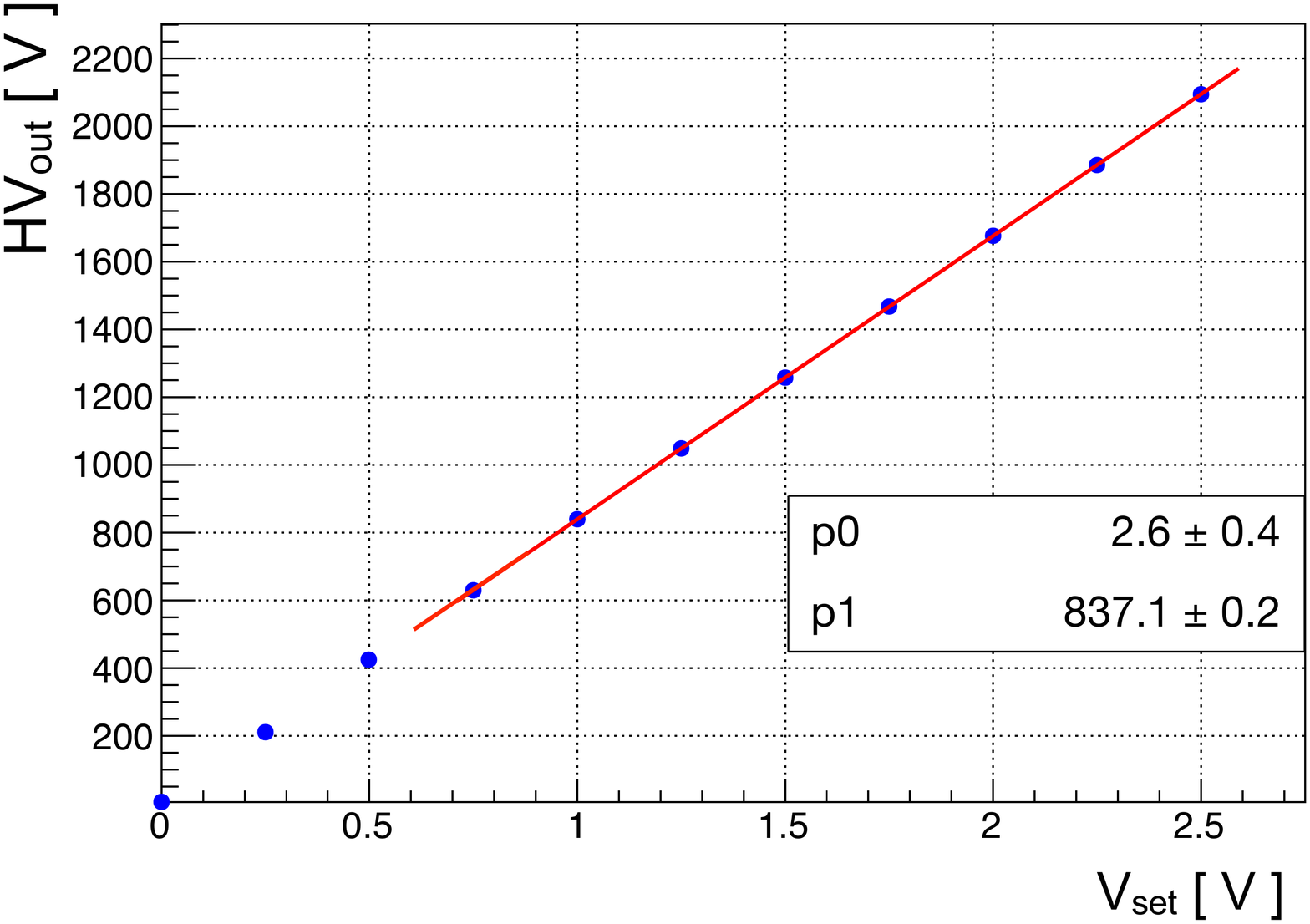}
\includegraphics[width=.49\textwidth]{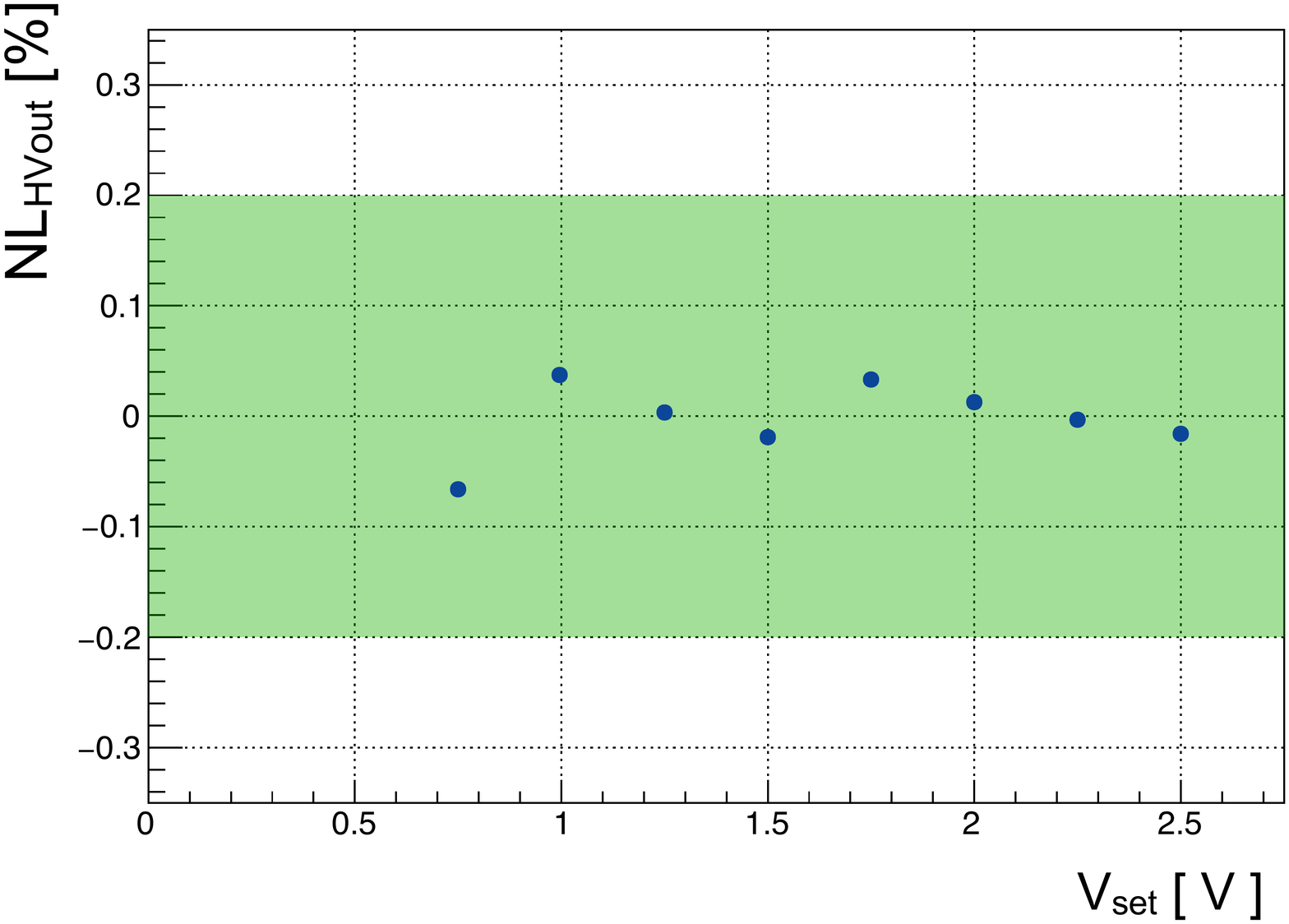}
\vspace{-0.25cm}
\caption{\label{fig:linearity} \textit{Left}: $\mathrm{HV_{out}}$ vs $\mathrm{V_{set}}$ for a single module. \textit{Right}: corresponding integral non linearity. The uncertainties over these quantities are below few \%, thus the error bars are hidden by the markers in both plots. The green area identifies acceptable values according to the requirement in table~\ref{table1}.}
\end{figure}

An example of the measured correlation between $\mathrm{HV_{out}}$ and $\mathrm{V_{set}}$ for one module is shown in the left panel of figure~\ref{fig:linearity}, where a linear fit for $\mathrm{HV_{out}}$ above 500~V is superimposed. 
In the right panel of the same figure, the behavior of the integral non linearity, defined as the relative difference between the measured parameter and the one estimated by the linear fit, is depicted. Small deviations from linearity, larger than the uncertainties but within the requirement in table~\ref{table1}, are intrinsic of each module when coupled with the testing system.

The $\mathrm{V_{set}}$ ramp is repeated three times, setting the HV module supply voltages to 3 different values: $\mathrm{V_{cc}}$ = 11.5~V, 12.0~V and 12.5~V. 
This test enables evaluation of the HV$_{\mathrm{out}}$ stability for small changes of the main supply voltage, an indeed conceivable situation when the HVPSs are operating in the field and supplied by solar panels and batteries.

\subsection{Uniformity measurements}
\label{subsec:uniformity}

\begin{figure}[tb]
\centering
\includegraphics[width=.49\textwidth]{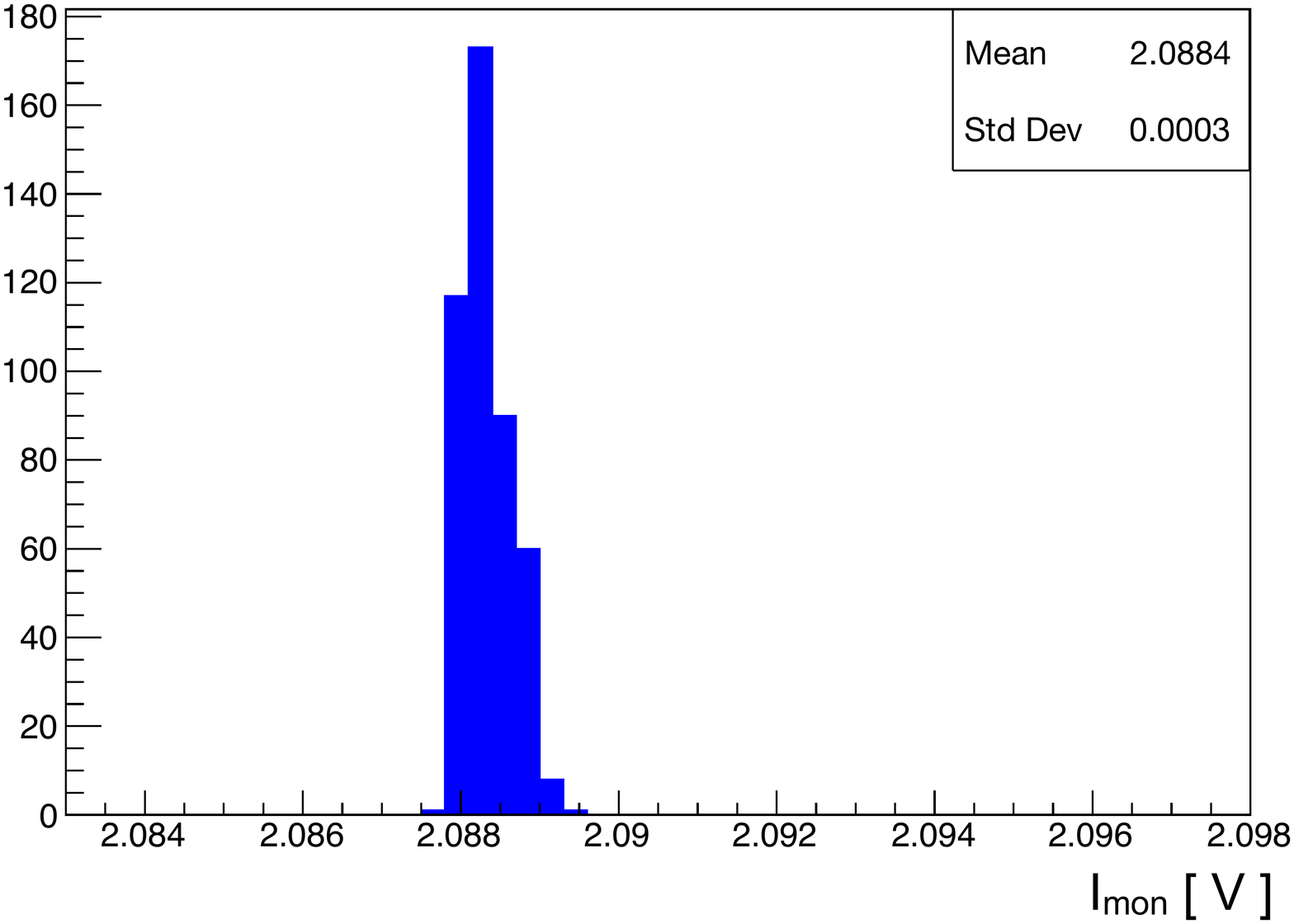}
\includegraphics[width=.49\textwidth]{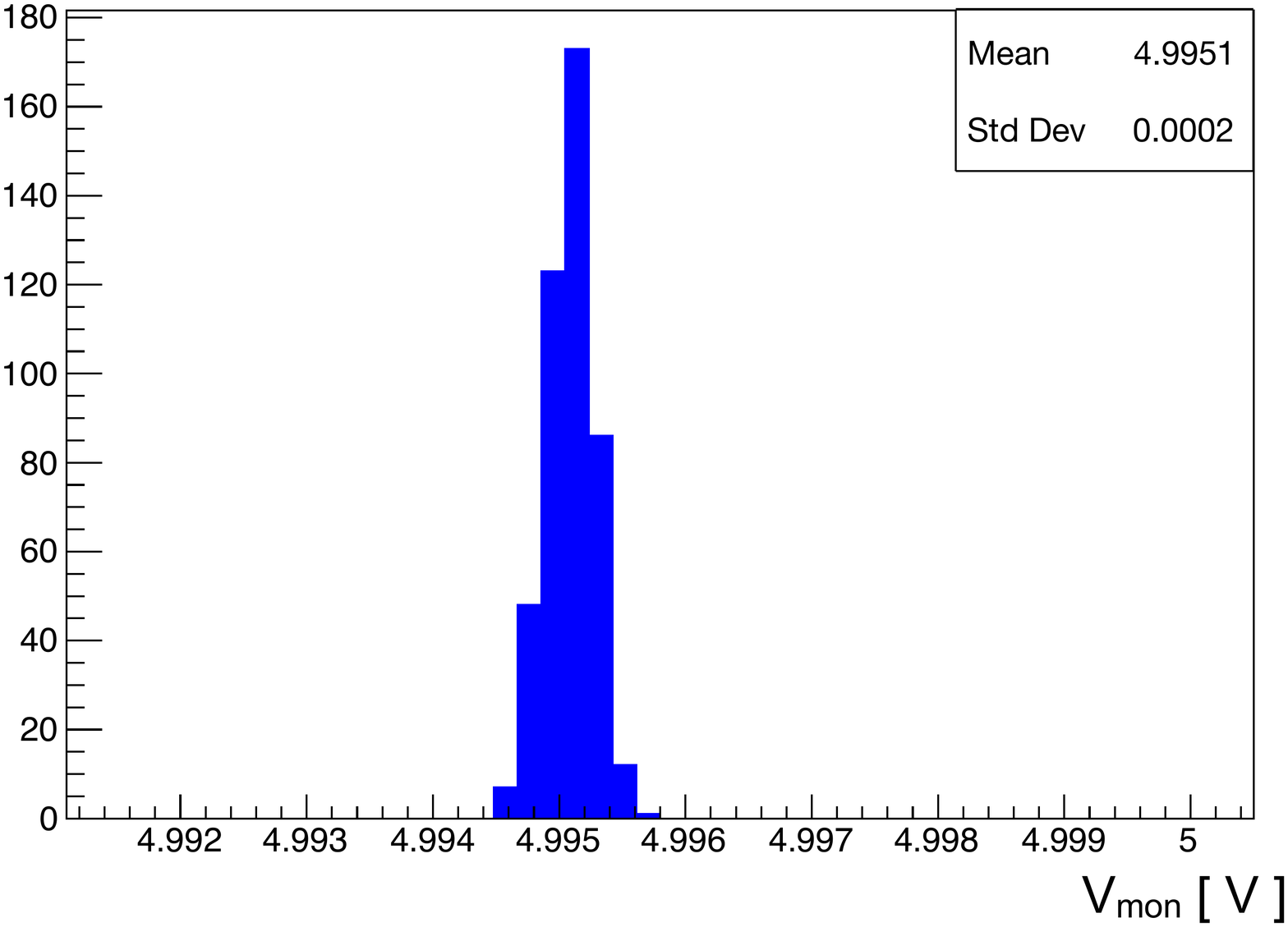}
\vspace{-0.4cm}
\caption{\label{fig:uniformity} \textit{Left}: $\mathrm{I_{mon}}$ at $\mathrm{V_{set}}=\mathrm{V_{set2}}$ = 2.5~V for a single module. \textit{Right}: same as left panel for $\mathrm{V_{mon}}$.}
\end{figure}

The uniformity measurement is used to study the HV$_{\mathrm{out}}$ response when a series of identical $\mathrm{V_{set}}$ commands are sent to the modules. The program consists of 45 cycles where the control voltage $\mathrm{V_{set}}$ changes alternatively between two different values ($\mathrm{V_{set1}}$= 1.25~V and $\mathrm{V_{set2}}$ = 2.5~V). For each step, all the monitoring parameters are read (with a ten-time repetition) and their values recorded to be analysed. 
The first fifteen cycles are used only to warm and stabilise the modules and are usually discarded for uniformity evaluation. An example of the test results for one module is presented in figure~\ref{fig:uniformity}: the two panels show the distribution of $\mathrm{I_{mon}}$ and of the corresponding $\mathrm{V_{mon}}$ for a single module in the case of $\mathrm{V_{set}} = \mathrm{V_{set2}}$.

\subsection{Thermal stress measurements}
\label{subsec:thermal}

\begin{figure}[tbp]
\centering
\includegraphics[width=.99\textwidth]{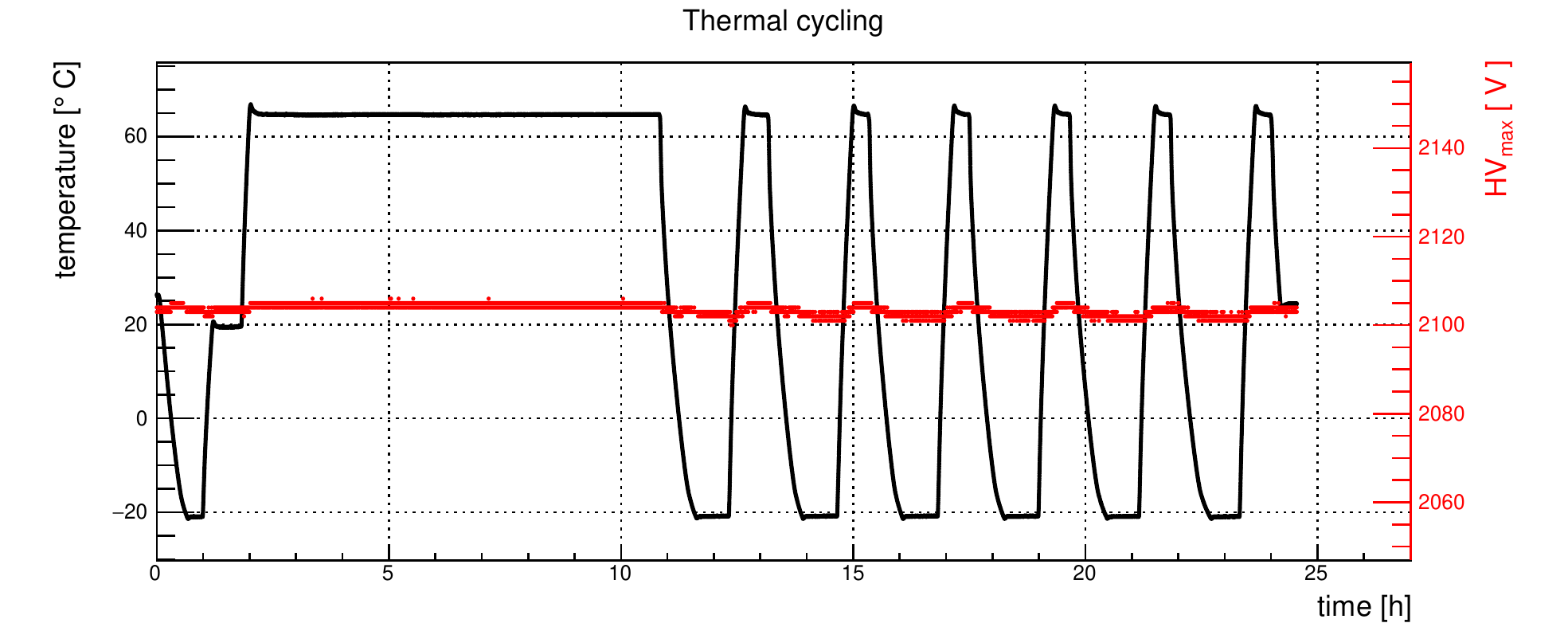}
\includegraphics[width=.99\textwidth]{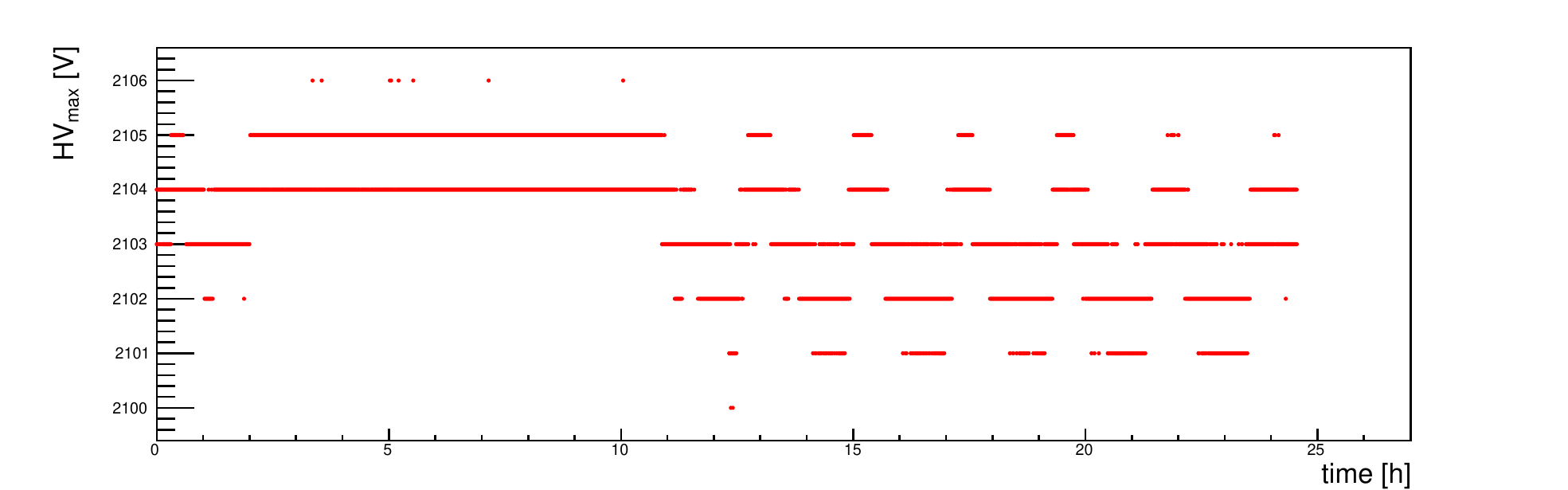}
\caption{\label{fig:temperatureCycle} \textit{Top}: temperature profile for the 24-hour cycle in the climatic chamber, with the corresponding measurement of $\mathrm{HV_{max}}$ for a single module in red. \textit{Bottom}: a zoomed version of the $\mathrm{HV_{max}}$ profile is shown to appreciate the small variations (order of 1\textperthousand).}
\end{figure}

As mentioned before, both the linearity and the uniformity measurements are performed at room temperature, the time required to study eight HVPSs in parallel being about 15 minutes. When the first two steps are completed, the eight modules under test are placed in the climatic chamber, where a 24-hour burn-in/thermal stress is performed following the temperature profile shown in the upper panel of figure~\ref{fig:temperatureCycle}, where the temperature ranges between -20$^\circ$C and +65$^\circ$C.
This test also assures the endurance of all the HVPS components to the continuous daily temperature variations.

The $\mathrm{V_{set}}$ command of all the modules is kept at the maximum value of 2.5~V for the full temperature cycle and all the other parameters are recorded each second.
In the lower panel of figure~\ref{fig:temperatureCycle}, the effect of the temperature variation on the output voltage of a module is shown.

\subsection{The test bench}
\label{subsec:test_bench}

Two identical test benches, able to manage up to 8 HVPSs in parallel, have been setup at the INFN-Torino and INFN-Catania laboratories.

\begin{figure}[t]
\centering
\includegraphics[width=.8\textwidth]{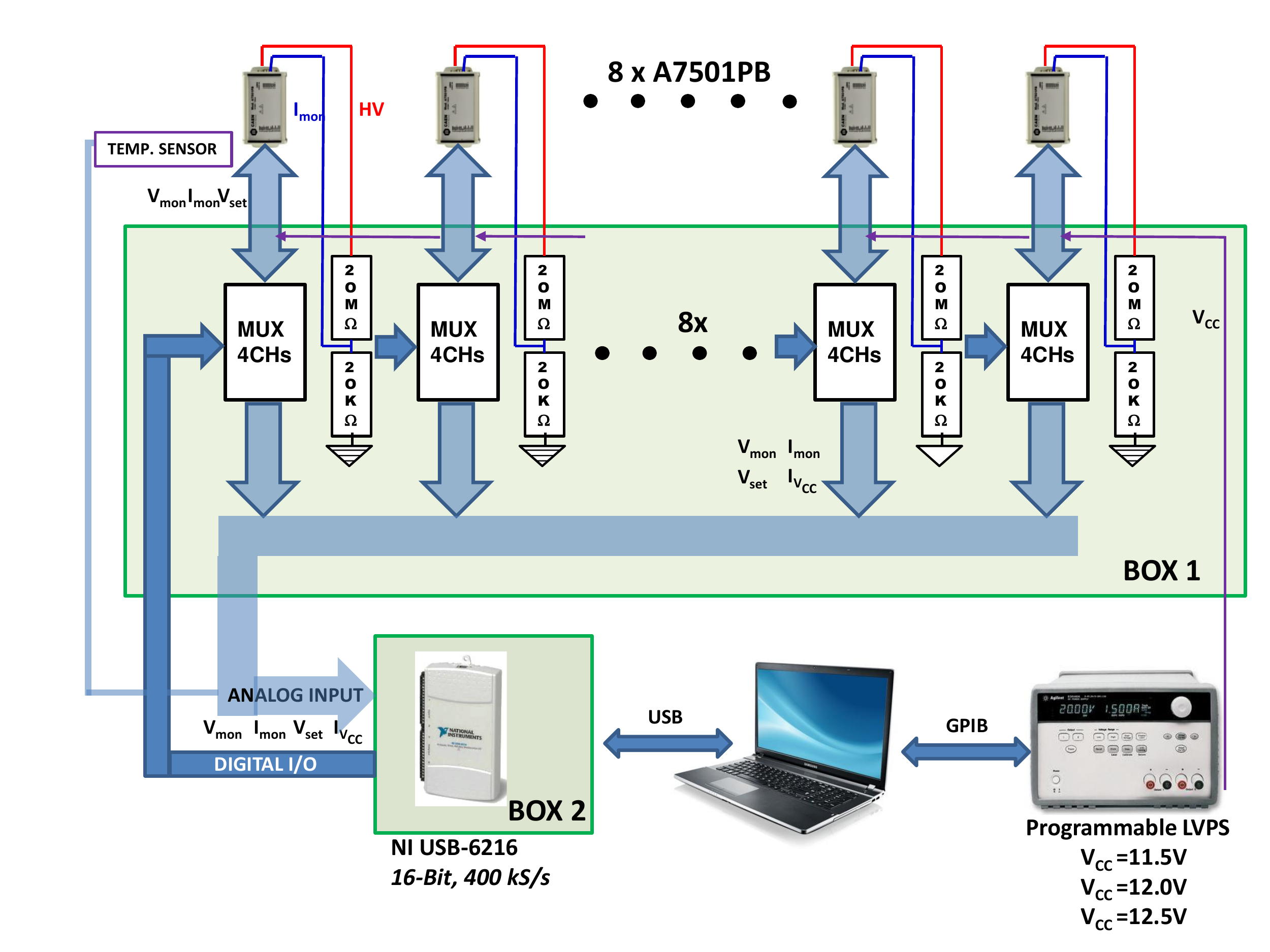}
\vspace{-0.5cm}
\caption{\label{fig:testBench} Sketch of the test bench circuitry (detailed description in the text).}
\vspace{0.5cm}
\end{figure}

\begin{figure}[t]
\centering
\includegraphics[trim= 0 230 0 330 , clip,  width=.44\textwidth]{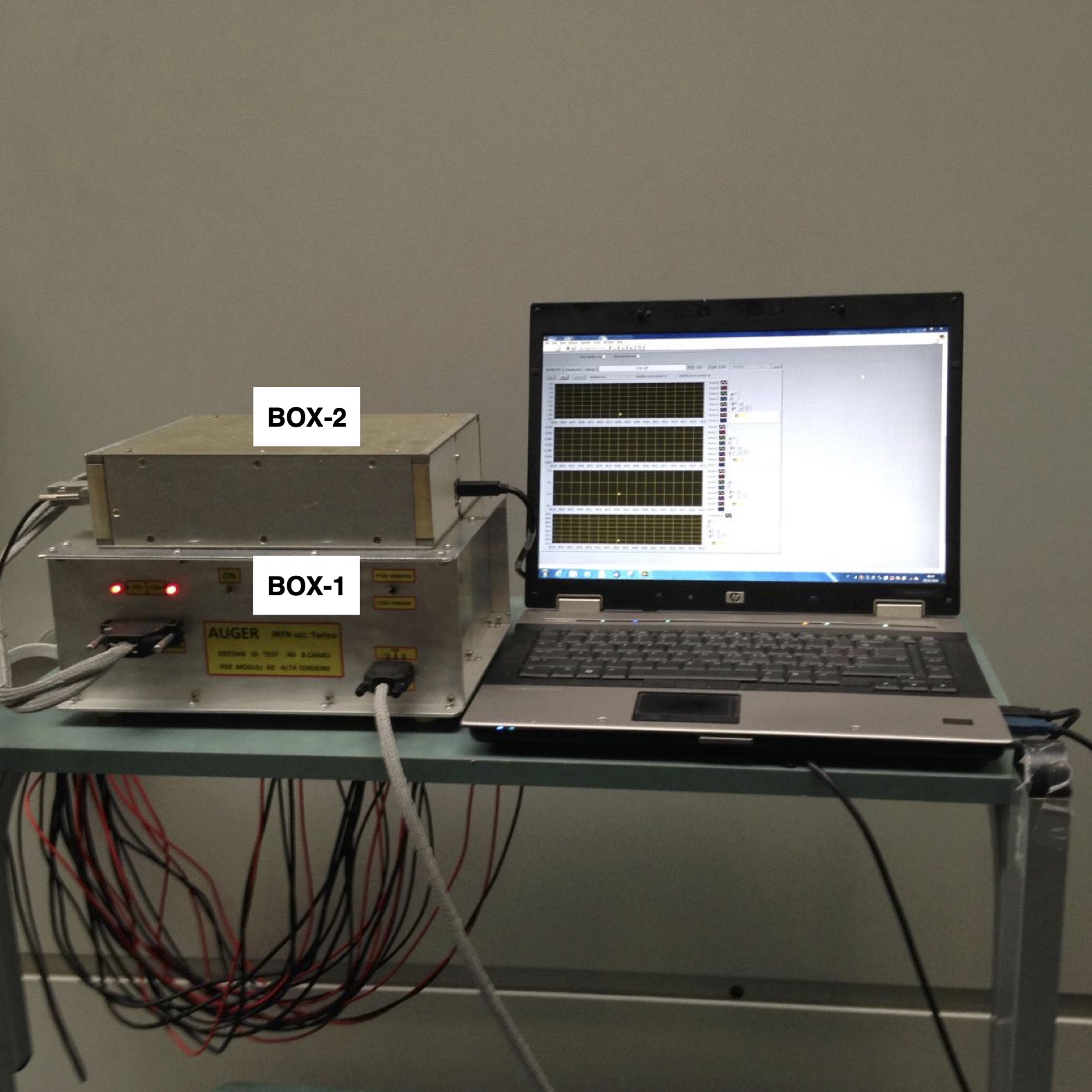}
\qquad
\includegraphics[width=.5\textwidth]{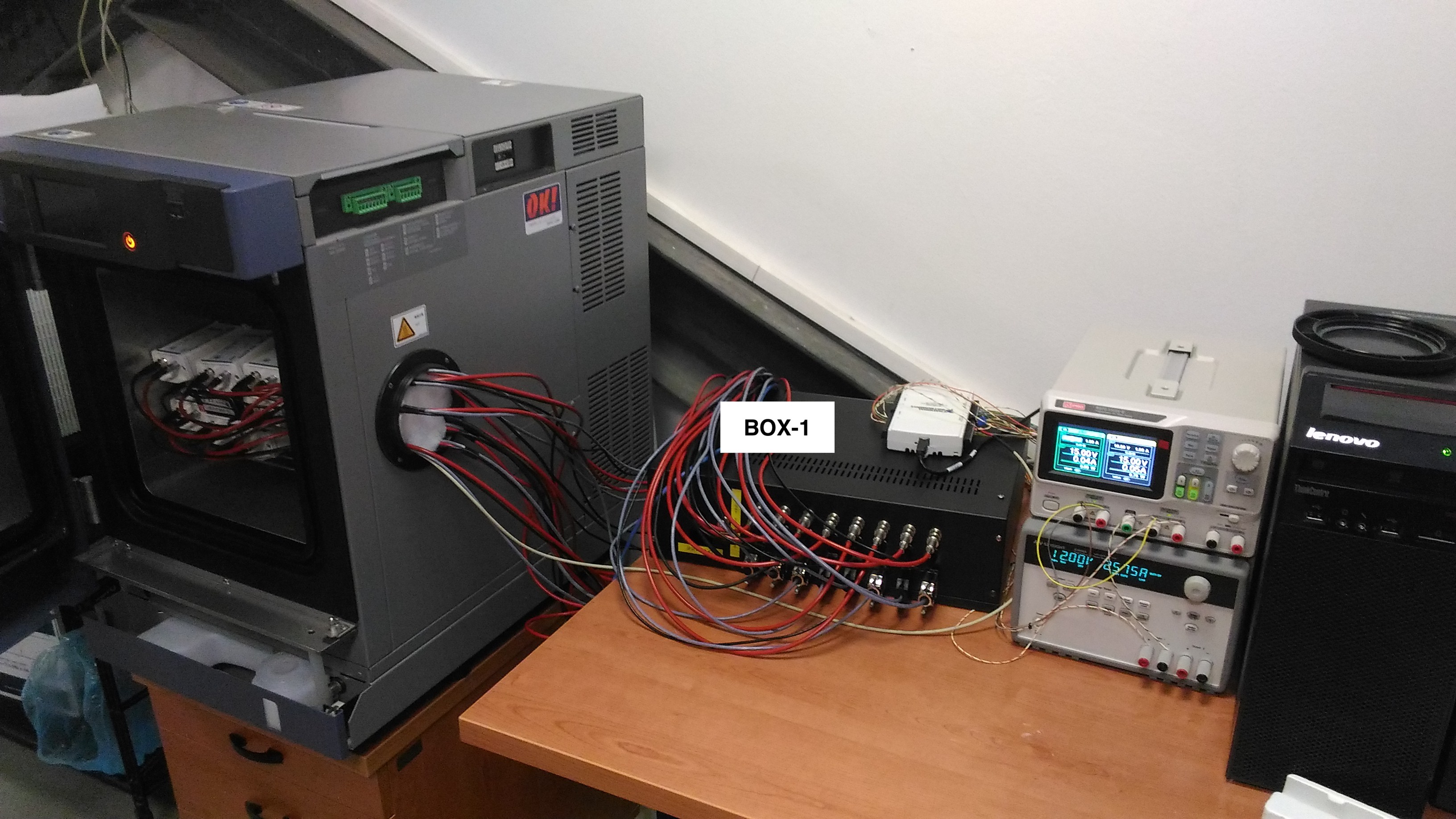}
\caption{\label{fig:setupPhotos} The test bench: the setup in Torino and the one in Catania in the left and right panels respectively.}
\end{figure}

The core element of the test system is a 400 kS/s National Instruments USB-6216 USB multi-function data acquisition (DAQ) with 16 analog inputs, two analog output and 32 digital I/O lines. Four of these lines are used to drive eight ADG452 analog switch ICs (one for each HVPS under test) and to route the parameters to be read ($\mathrm{I_{mon}}$, $\mathrm{V_{mon}}$, $\mathrm{I_{V_{cc}}}$) to the USB-6216 analog input associated to the module. 
The resistive loads required by the HV modules, eight 20~M$\Omega$ resistors, together with the low voltage HV monitor outputs and the circuits devoted to the measurement of the absorption current $\mathrm{I_{V_{cc}}}$ are all located outside the climatic chamber in the same box hosting also the ADG452 switches (figure~\ref{fig:testBench}, BOX-1). The USB card has been installed, alone, in a smaller separate unit to simplify the internal interconnections (figure~\ref{fig:testBench}, BOX-2).
In figure~\ref{fig:setupPhotos}, two photographs of the full assembly are shown.

The full test procedure is automated and controlled by a LabView code locally running on a laptop; it also controls, via GPIB, the low voltage power unit used to supply the HVPSs at three different values.

\newpage
\section{Results of the validation}
\label{sec:results}

A total of 1650 HVPSs modules have been tested and delivered to the Pierre Auger Observatory. 
As discussed in the introduction, the following analysis only includes the 1560 modules with final design, not considering the 90 employed in the pre-production array.

All the modules were found to be working after the first preliminary thermal stress, confirming the accuracy and reliability of their assembly. 
The validation tests performed at room temperature were successful and compatible with the measurements provided by CAEN for each module.
On the other hand, less than 1\% of the HVPSs were found to be outside of the required values after the final thermal stress phase and were sent back to the manufacturer for substitution. This kind of failures couldn't be caught during the tests reported by CAEN since those were performed at room temperature only.

\begin{figure}[hbp]
\centering 
\includegraphics[width=.49\textwidth]{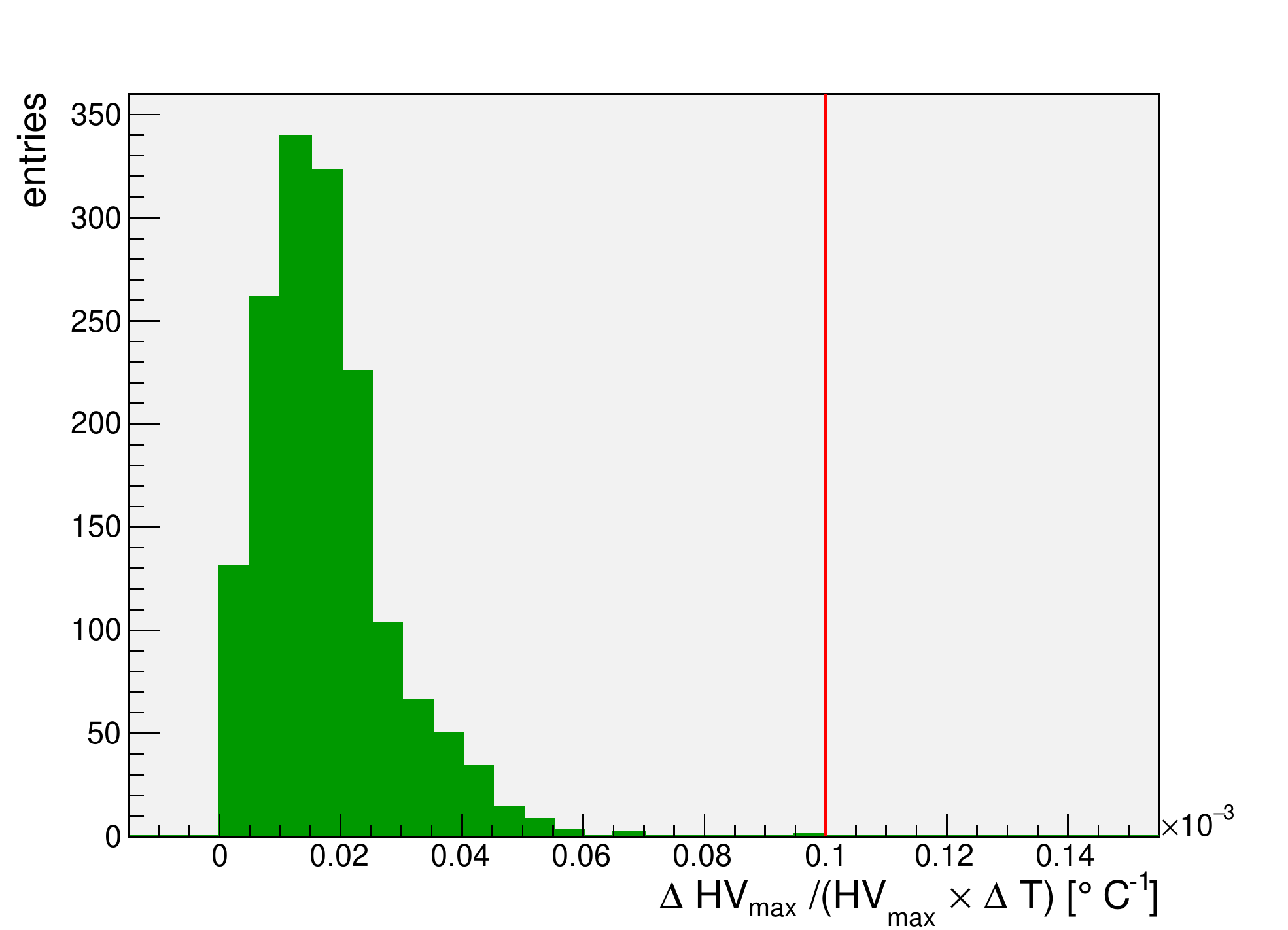}
\includegraphics[width=.49\textwidth]{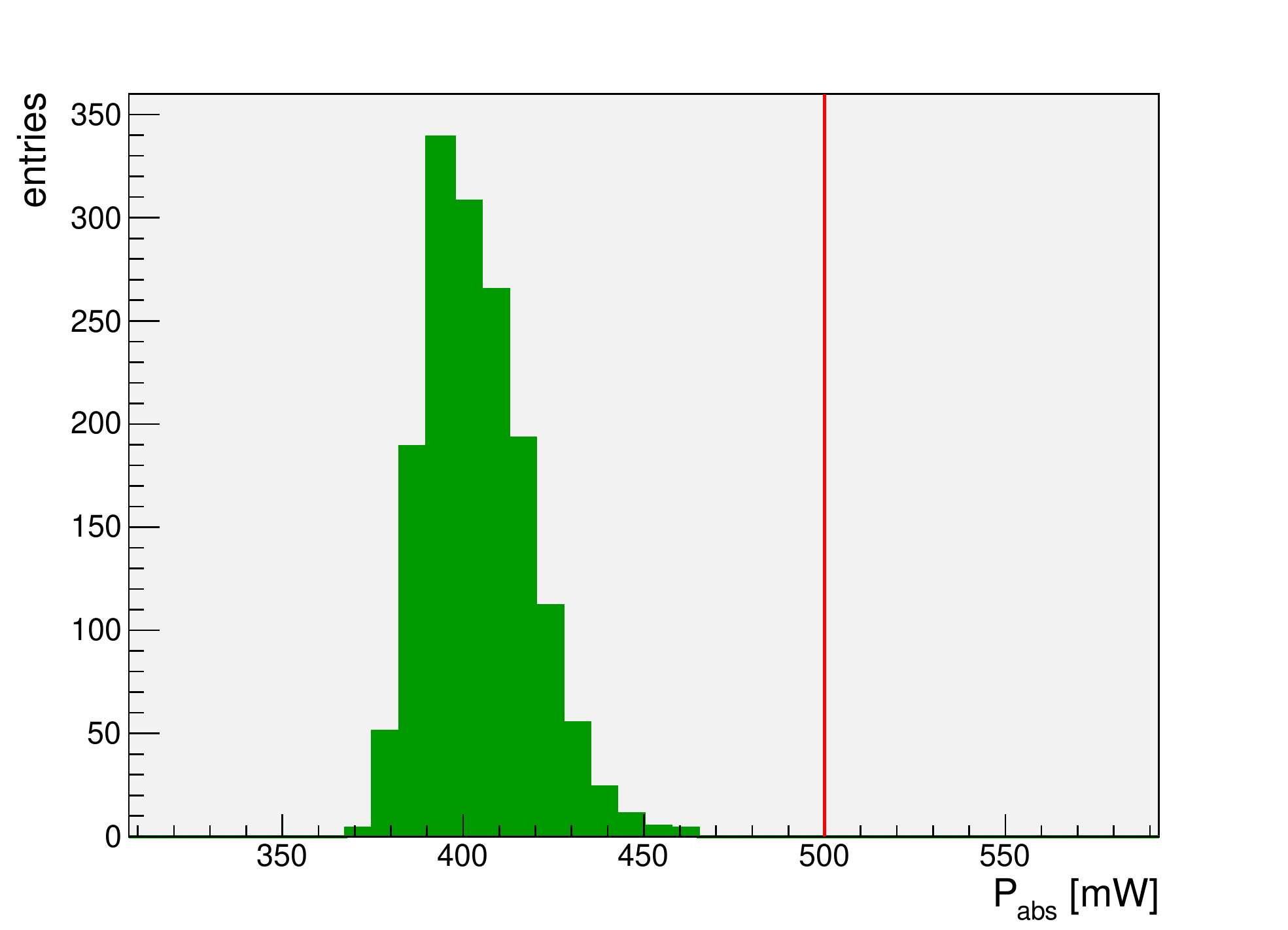}
\caption{\label{fig:ThermalStability_Power} \textit{Left}: distribution of the thermal stability parameter, evaluated as $\Delta \mathrm{HV_{max}}/(\mathrm{HV_{max}} \times \Delta \mathrm{T})$. \textit{Right}: distribution of the power absorption value ($\mathrm{P_{abs}}$). The red lines indicate the maximum acceptable values according to table~\ref{table1}.}
\end{figure}

In figure~\ref{fig:ThermalStability_Power} the distributions for the thermal stability (left), defined as $\Delta \mathrm{HV_{max}}/(\mathrm{HV_{max}} \times \Delta \mathrm{T})$, and for the power absorption $\mathrm{P_{abs}}$ (right) are shown.\footnote{The tested variables are defined in table~\ref{table1}.}
In the former, the majority of the modules presents a coefficient lower than $5\times 10^{-4}/^\circ$C that is half of the required specification, ensuring a good stability even in extreme conditions. In the latter, the bulk of values is peaked around 400~mW @ $\mathrm{HV_{max}}$ and 20~M$\Omega$ load, well below the threshold suitable for low consumption conditions.
Moreover, in the field the HV required by the SPMT is around 1000~V and the resistive load is 68~M$\Omega$, resulting in a effective power consumption $\sim$150~mW.

\begin{figure}[htbp]
\centering 
\includegraphics[width=.45\textwidth]{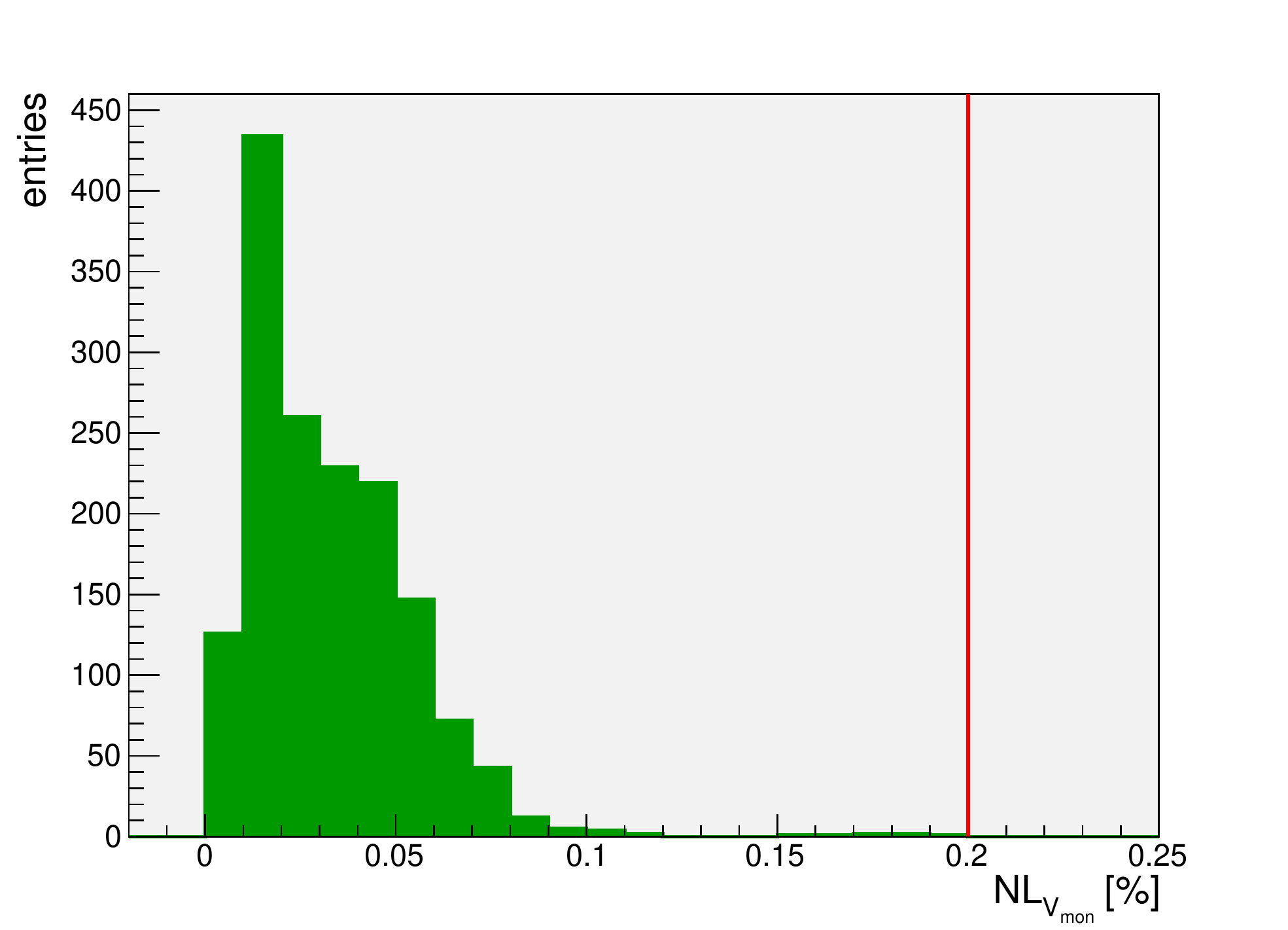}
\qquad
\includegraphics[width=.45\textwidth]{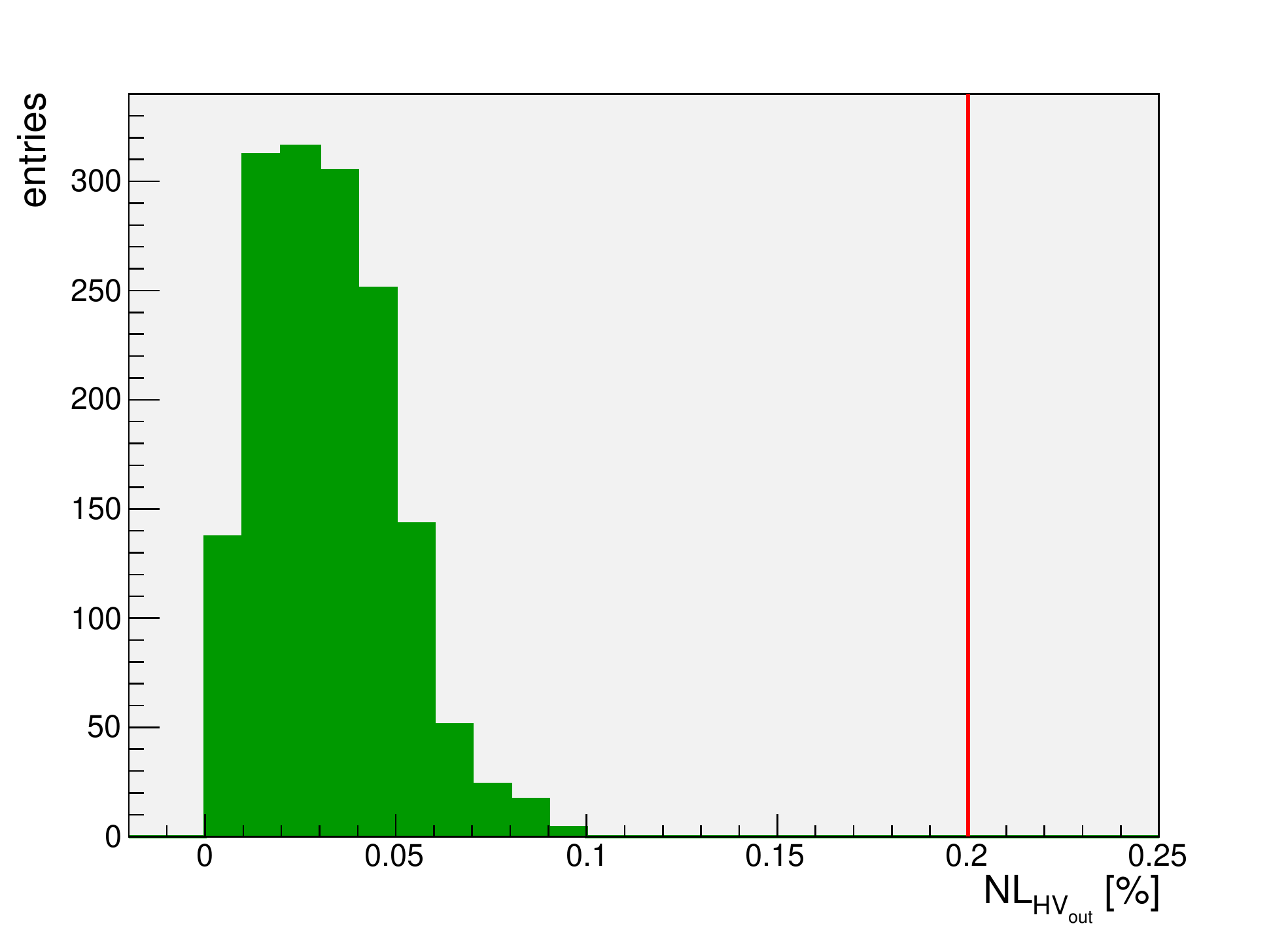}
\qquad
\includegraphics[width=.45\textwidth]{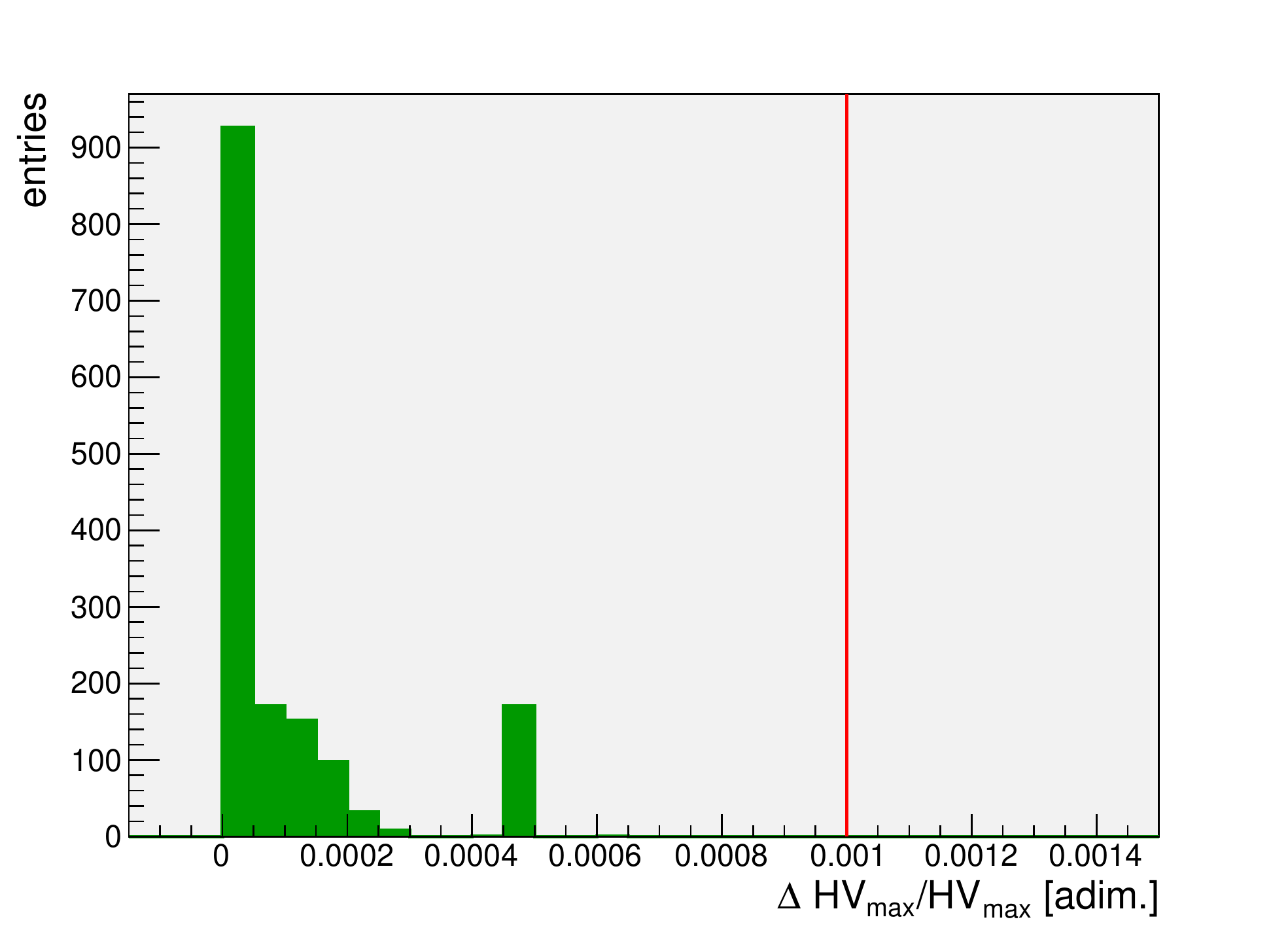}
\qquad
\includegraphics[width=.45\textwidth]{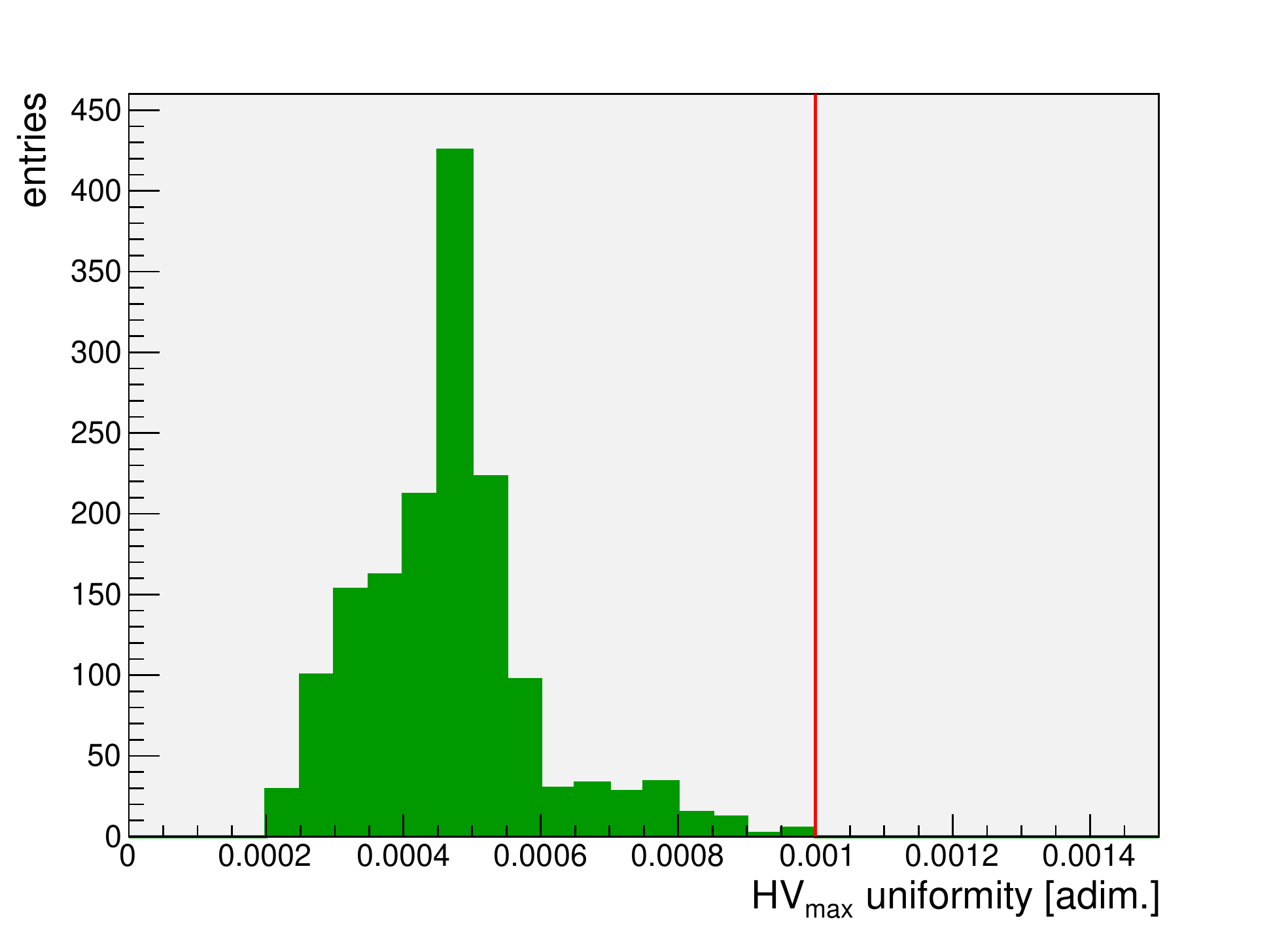}
\qquad
\includegraphics[width=.45\textwidth]{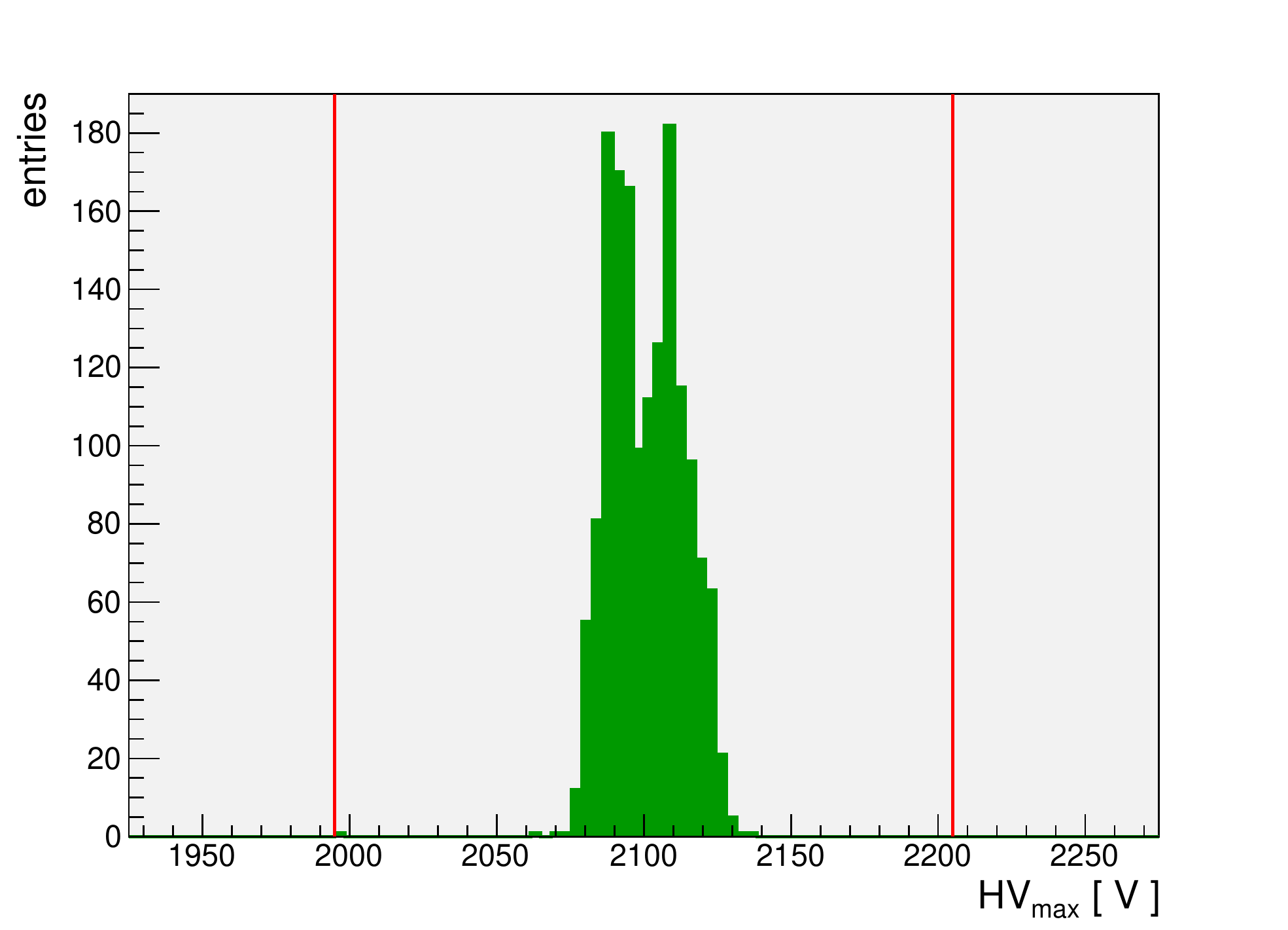}
\caption{\label{fig:otherDistrib}\textit{Top left}: distribution of the integral non linearity for $\mathrm{V_{mon}}$ vs $\mathrm{HV_{out}}$ ($\mathrm{NL_{V_{mon}}}$). \textit{Top right}: distribution of the integral non linearity for $\mathrm{HV_{out}}$ vs $\mathrm{V_{set}}$ ($\mathrm{NL_{HV_{out}}}$). \textit{Middle left}: stability of $\mathrm{HV_{max}}$ vs $\mathrm{V_{cc}}$ ($\Delta \mathrm{HV_{max}}/\mathrm{HV_{max}}$). \textit{Middle right}: uniformity of $\mathrm{HV_{max}}$. \textit{Bottom}: distribution of the maximum output voltage $\mathrm{HV_{max}}$ at room temperature. The red lines indicates the maximum acceptable values according to table~\ref{table1}.}
\end{figure}

The overall distribution for the other tests performed at room temperature (non-linearity, stability, uniformity) plus the effective maximum output voltage measured for each module are reported in figure~\ref{fig:otherDistrib}.
It can be observed that all the modules fulfil the requirements, represented by the red lines according to the values in table~\ref{table1}. Peculiarities in each histogram are explained by unavoidable small differences in the two test benches and in the 8 independent channels of each setup, and moreover considering that the data taking was carried on during a 2 year long period.

\newpage

\section{Conclusions}
\label{sec:conclusions}

To extend the dynamic range of the measurement in the Surface Detector stations of the Pierre Auger Observatory, a small PMT of 1-inch photocatode diameter is being added to the WCD. 
New HVPS modules have been designed and integrated in the upgraded unified board to supply the SPMT. 
An accurate validation procedure needed to ensure that they can survive the harsh environment in the field has been defined and brought on using two identical facilities developed in the laboratory.
About 1600 CAEN A7501B HVPSs have been successfully validated and are currently being installed in the experimental site.

\newpage
\appendix

\section{Extract of the documentation provided by CAEN for one HVPS module}
\label{appendix:a}
\begin{figure}[htbp]
\centering
\includegraphics[width=.75\textwidth]{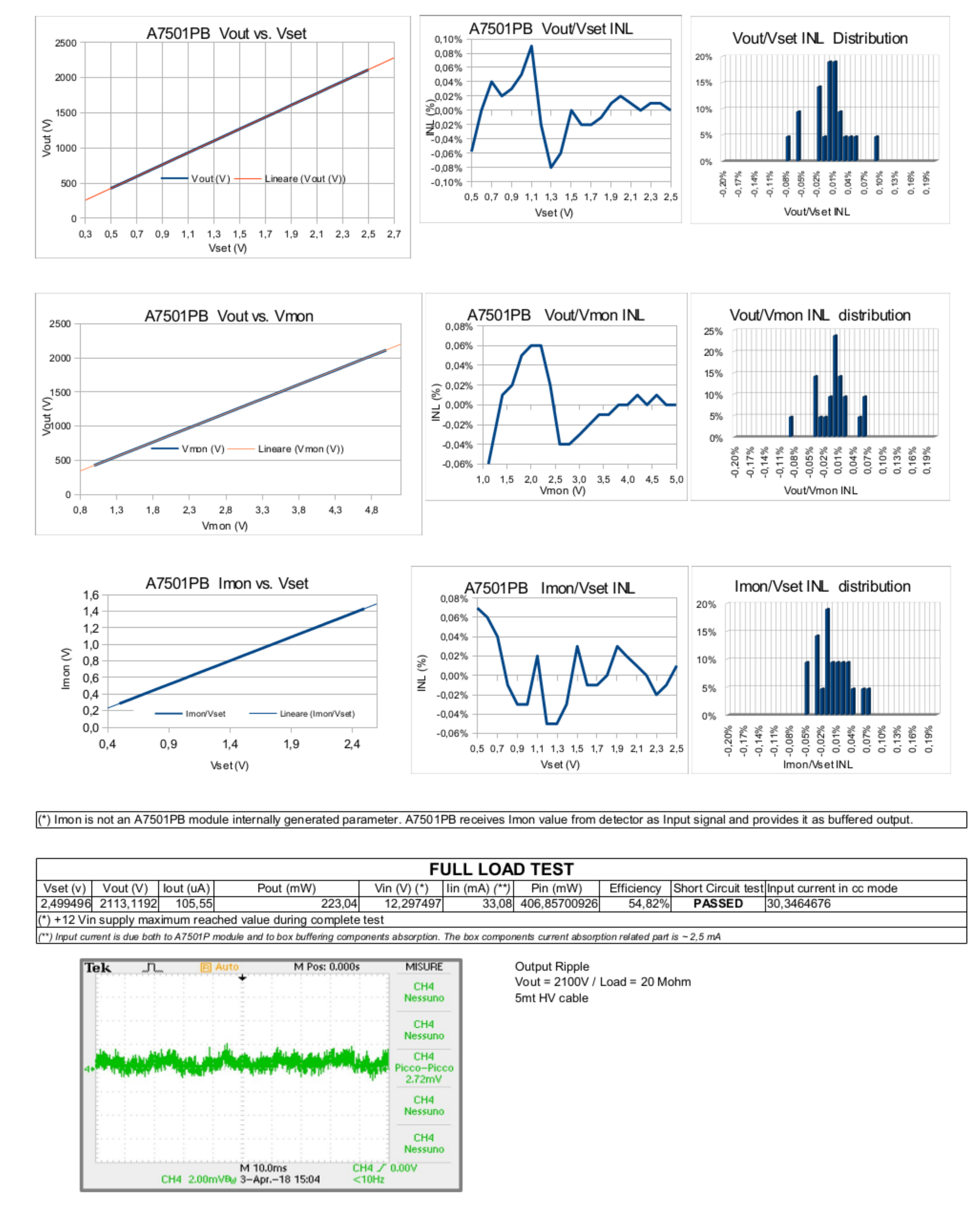}
\caption{\label{fig:estratto_testCAEN} Results of the tests performed by the manufacturer for one module at room temperature. 
\textit{Upper plots}: output high voltage (here called $\mathrm{V_{out}}$ instead of $\mathrm{HV_{out}}$) vs $\mathrm{V_{set}}$ (\textit{left}); deviation from linearity of $\mathrm{V_{out}}$ vs $\mathrm{V_{set}}$ (\textit{center}) and corresponding histogram (\textit{right}). 
\textit{Middle plots}: output high voltage $\mathrm{V_{out}}$ vs $\mathrm{V_{mon}}$ (\textit{left}); deviation from linearity of $\mathrm{V_{out}}$ vs $\mathrm{V_{mon}}$ (\textit{center}) and corresponding histogram (\textit{right}). 
\textit{Bottom plots}: $\mathrm{I_{mon}}$ vs $\mathrm{V_{set}}$ (\textit{left}); deviation from linearity of $\mathrm{I_{mon}}$ vs $\mathrm{V_{set}}$ (\textit{center}) and corresponding histogram (\textit{right}). 
In the lower part, a table with the parameter values measured at $\mathrm{V_{set}} = 2.5$~V and the ripple amplitude.}
\end{figure}

The manufacturer provided a detailed documentation for the tests (only at room temperature) performed on each module. 
An example of such reports is shown in figure~\ref{fig:estratto_testCAEN}. 

Similarly to the linearity test described in section~\ref{subsec:linearity}, the values of the output high voltage (here called $\mathrm{V_{out}}$ instead of $\mathrm{HV_{out}}$), output voltage monitor $\mathrm{V_{mon}}$ and output current monitor $\mathrm{I_{mon}}$ are stored for different values of the external voltage setting $\mathrm{V_{set}}$ to study the deviations from linearity.
Moreover, a table with several characteristic parameters @ full load (i.e.~$\mathrm{V_{set}} = 2.5$~V) is reported in the lower part, together with the measurement of the ripple amplitude.

\vspace{0.2cm}

%



\end{document}